\documentclass[aps,prl,twocolumn,superscriptaddress,amsmath,amssymb,showpacs,nofootinbib,reprint]{revtex4-1}

\usepackage{graphicx}
\usepackage{dcolumn}
\usepackage{bm}
\usepackage{color}

\begin{document}

\title{Tunneling Conductance and Surface States Transition in Superconducting Topological Insulators}

\author{Ai Yamakage}
\author{Keiji Yada}
\affiliation{Department of Applied Physics, Nagoya University, 
Nagoya 464-8603, Japan}

\author{Masatoshi Sato}
\affiliation{Institute for Solid State Physics, University of Tokyo, 
Chiba 277-8581, Japan}

\author{Yukio Tanaka}
\affiliation{Department of Applied Physics, Nagoya University, 
Nagoya 464-8603, Japan}

\date{\today}

\begin{abstract}

We develop a theory of the tunneling spectroscopy for
superconducting topological insulators (STIs), where the surface Andreev bound
states (SABSs) appear as helical Majorana fermions. 
Based on the symmetry and topological nature of parent
 topological insulators, 
we find that the SABSs in the STIs have a profound structural
transition in the energy dispersions.
The transition results in a variety of Majorana fermions, by tuning 
the chemical potential and the effective mass of the energy band. 
We clarify that Majorana fermions in the vicinity of the
 transitions give rise  to robust zero bias peaks in the tunneling
 conductance between normal metal/STI junctions.

\end{abstract}


\maketitle

Topological superconductors (TSCs) are a new state of matter 
\cite{TSN, QZ,Schny_B08} that characterized by non-zero topological
numbers of the bulk wave functions.
They support topologically protected gapless surface Andreev bound states
(SABSs), and  the superconductivity infers that 
the gapless SABSs are their own
antiparticles, thus Majorana fermions \cite{Wil_NP09}.
The realization of TSC and Majorana fermions in condensed matter physics
is of particular interest because of their novelty as well as the possible
application for quantum devices 
\cite{SF09,STF09,STF10,SF10,SLTS10,Alicea10,LSD10,ORO10,LSS11,AOROF11,Fu1,FK09,ANB09,LLN09,TYN09,Linder10a,yamakage}.

The recent discovered superconductor Cu$_x$Bi$_2$Se$_3$ 
\cite{Hor_L10,Wray_NP10,MKR_L11,MKR2} is an intriguing candidate of the TSC 
because it is associated with another new state of matter, topological
insulator:
The parent material Bi$_2$Se$_3$ is originally a topological insulator with
topologically protected gapless Dirac 
fermions on its surface. 
With intercalating Cu, the superconductivity appears. 
On the theoretical side, 
it has been expected 
that Cu$_x$Bi$_2$Se$_3$ 
is a TSC by the
Fermi surface criterion \cite{Sato_B09,FB_L10,Sato_B10}, and
possible SABSs specific to this material have been studied
\cite{HaoLee_B11, SasakiPRL,Hsieh-Fu,hsieh11}.
Recently, a point contact spectroscopy experiment on this material has
been done, and reported a pronounced zero-bias conductance peak
(ZBCP) \cite{SasakiPRL}.
With careful analysis excluding other mechanisms, it has been concluded
that the ZBCP is intrinsic 
and signifies
unconventional superconductivity \cite{SasakiPRL}. 
Moreover, similar ZBCPs have been observed by other 
groups independently \cite{Kirzhner,Koren,FanYang}.

Motivated by this finding, we develop in this letter a general theory of
Majorana fermions in superconducting topological insulators (STIs)
and their relation to the tunneling conductance. 
Up to this time, the relation between SABSs and
the tunneling conductance has been well understood 
in quasi two-dimensional superconductors \cite{TSN}: 
(1) If the SABS has a flat band dispersion 
as a function of the momentum parallel to 
the surface, $k_{y}$, the resulting 
line shape of conductance always has a sharp ZBCP 
as realized in high-$T_{\rm c}$ cuprate \cite{TSN,TK95}. 
(2) If the SABS has a linear  dispersion 
as a function of $k_{y}$, 
the resulting  line shape of conductance has a broad peak as 
observed in Sr$_{2}$RuO$_{4}$ \cite{Sigrist,YTK97,Kashiwaya11}. 
On the other hand, in three-dimensional superconductors, little is known
about the relation between SABSs and the tunneling
conductance.
The only exception is a study on the superconducting analog of
superfluid $^3$He B phase \cite{Asano}.
Like Cu$_x$Bi$_2$Se$_3$, it is a three dimensional TSC supporting
helical Majorana fermions on its surface
\cite{Qi,Schnyder,Nagato,Volovik1,Volovik2,Tsutsumi}. 
However, the resulting tunneling conductance always shows a double-peak
structure and never shows a ZBCP \cite{Asano}. 
%
Therefore, 
in order to pursue the origin of the observed ZBCPs in STIs, one needs to
develop a theory of the 
tunneling conductance for STIs.

In this letter, we study the tunneling 
spectroscopy and underlying SABSs in STIs. 
Based on symmetry and topological nature of parent topological
insulators, 
it is shown that SABSs in STIs generally have a profound structural transition 
of the energy dispersion (Fig.\ref{edge}).
The transition results in a variety of helical
Majorana fermions in SABSs, which we call cone, caldera, ridge and valley.
%
We clarify that the transition naturally explains robustness of the
ZBCP in STIs.
From explicit calculation, it is found that
the tunneling conductances between normal metal/STI junctions ubiquitously
support ZBCPs near the transition.
These features are proper to STIs and distinct from a simple three-dimensional 
TSC mentioned above.
Our findings strongly support that the observed ZBCPs in
Refs.\cite{SasakiPRL,Kirzhner,Koren,FanYang} are originated from a helical
Majorana fermion in 
STIs, and they give a firm evidence of their topological superconductivity. 
Our results are summarized in Table \ref{gap}.

First let us briefly review basic properties of parent topological insulators.
For concreteness, we consider the following $k\cdot p$ Hamiltonian to describe
the topological insulators,
\begin{eqnarray}
 H_{\rm TI}(\bm k) = m \sigma_x + v_z k_z \sigma_y +v \sigma_z (k_x s_y
 - k_y s_x),
 \nonumber\\
 m = m_0 + m_1 k_z^2 + m_2 (k_x^2+k_y^2), \,(m_1m_2>0),
\end{eqnarray}
where $s_{\mu}$ and $\sigma_{\mu}$ are the Pauli matrices in 
spin and orbital spaces, respectively.
In addition to the time-reversal symmetry, we have
assumed the mirror symmetry 
$\mathcal M_i H_{\rm TI} \mathcal M_i^\dag 
= H_{\rm TI}|_{k_i \to -k_i}$
with 
$\mathcal M_i = s_i$, ($i=x,y$) and the inversion symmetry. 
Although
$H_{\rm TI}$ in the above is axial symmetric along the $z$-axis,
even if one adds
higher order terms of $k_i$ ($i=x,y$)
like the warping terms \cite{fu09b},
our results do not change qualitatively.
The topological phase of this system is classified by the $\mathbb Z_2$
invariant, $(-1)^{\nu}={\rm sgn}(m_0m_1)$, and when $\mathbb Z_2$ non-trivial $(m_0m_1<0)$, 
the system becomes a topological insulator. 
On the surface perpendicular to $z$-axis, 
it supports the topologically protected Dirac fermion.
%

\begin{table}
\begin{tabular}{c|c|c|c}
\hline\hline
& \multicolumn{2}{c|}{STI} & BW
\\
\hline
gap & full & nodal & iso
\\
SABS & cone/caldera & ridge/valley & cone
\\
conductance & DP/ZBP & ZBP & DP
\\
\hline\hline
\end{tabular}
 \caption{
Momentum-independent odd-parity paring symmetries 
in STI. 
As a comparison, pairing symmetry in  BW phase of superfluid $^{3}$He is 
shown. 
The energy spectrum has  full gap, nodal or isotropic (iso) full gap.
In cases of low and intermediate transmissivity of 
normal metal/STI junctions,
 the line shapes of tunneling conductances show double peak (DP) and zero bias peak (ZBP), respectively.  (see Fig.\ref{conductance} and the corresponding discussions in the text). 
}
\label{gap}
 \end{table}

Now consider the corresponding STIs. 
The STIs are described by the Bogoliubov-de Gennes (BdG) Hamiltonian in
the Nambu representation
$(\psi_{\sigma \uparrow}, \psi_{\sigma \downarrow},
-\psi^{\dagger}_{\sigma \downarrow}, \psi^{\dagger}_{\sigma \uparrow})$, 
\begin{eqnarray}
 H_{\rm STI}(\bm k) = (H_{\rm TI}(\bm k) - \mu) \tau_z + \hat\Delta \tau_x,
 \label{hamiltonian}
\end{eqnarray}
where $\mu$ is the chemical potential, $\tau_{\mu}$ is the Pauli
matrix in the particle-hole (Nambu) space, and
$\hat \Delta$ is a $4 \times 4$ matrix denoting the gap function.
For simplicity, we assume that $\hat \Delta$ is
a constant matrix, which is generally realized
when the pairing interaction is short-range and attractive.
Because of the Fermi-Dirac statistics of electrons, $\hat{\Delta}$ satisfies
$\hat{\Delta}=s_y\hat{\Delta}^{T}s_y$, thus
there are six independent pairings, $(\Delta, \Delta\sigma_x,
\Delta\sigma_z, \Delta\sigma_ys_x,\Delta\sigma_ys_y,\Delta\sigma_ys_z)$
($\Delta$ is independent of $\bm k$). 
For each independent pairing,
we consider SABS on the surface normal to the $z$-axis.

In order to solve the SABS, we consider the semi-infinite
STI ($z>0$) with a flat surface at $z=0$.
The wave function in this system is given by
\begin{align}
\psi_{\rm STI}(z>0) =
\sum_{I}
t_I u_I e^{iq_Iz}e^{ik_x x} e^{ik_y y},
\label{psi}
\end{align}
where $q_I$ $(I=1,\cdots,8)$ is a solution of $E=E(k_x,k_y,q_I)$ with
$E({\bm k})$ being an eigenvalue of Eq.(\ref{hamiltonian}), 
and $u_I(k_x,k_y,q_I)$ is the corresponding eigenvector.
Among the eigenvectors, $\psi_{\rm STI}(z)$ consists of
those with $E(k_x,k_y,q_I)/\partial q_I > 0$ or ${\rm Im} q_I > 0$,
where the former denotes up-going states and the latter denotes 
localized states 
in the vicinity of $z=0$.
Postulating the boundary condition as
$\psi_{\rm STI}(z=0) = 0$, we can determine the
coefficients $t_I$ and obtain the SABS.

\begin{figure}
\centering
\includegraphics[scale=0.7]{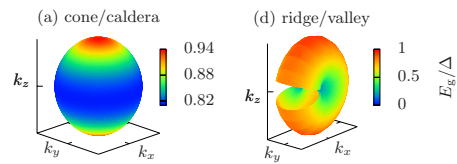}
\caption{
(color online)
Polar plots of the bulk superconducting gap $E_{\rm g}$ for full (a) and nodal (b) gaps, where cone/caldera and ridge/valley SABSs are realized, respectively.
It is not plotted in a certain region for the cases (b), for visibility.
}
\label{gap1-4}
\end{figure}

\begin{figure}
\centering
\includegraphics[scale=0.8]{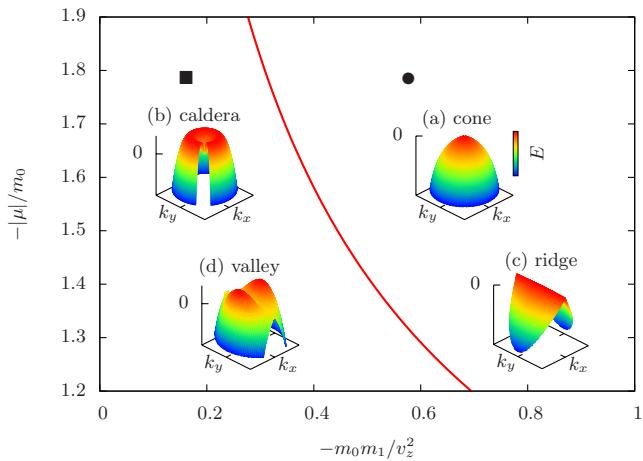}
\caption{
(color online)
Structural transition of SABSs in STIs. 
Cone (ridge) and caldera (valley) SABSs are shown in
the insets. 
The position of the circle (square) symbol corresponds to the parameters used in the calculations of tunneling conductances with $m_1=20.18$ eV\AA \ ($m_1=5.66$ eV\AA), where a cone (caldera) or ridge (valley)-SABS is realized.
}
\label{edge}
\end{figure}

We find that 
among the six pairings mentioned above,   
only the three $(\Delta\sigma_ys_x,
\Delta\sigma_ys_y,\Delta\sigma_ys_z)$ support well-localized gapless
SABSs on the surface at $z=0$.
We notice that all of them are odd-parity pairings, $P\Delta\sigma_y
s_{\mu}P^{\dagger} =-\Delta\sigma_ys_{\mu}$, ($P=\sigma_x$), and  
the existence of the SABSs is consistent with the Fermi surface
criterion for odd parity TSCs \cite{Sato_B10}.  
Furthermore, 
they all are odd under (at least) one of the mirror
symmetries, 
$\mathcal M_i \Delta \sigma_y s_{\mu} \mathcal M_i^\dag = - \Delta \sigma_y s_{\mu}$ for $i \ne \mu$.
As illustrated in Fig.\ref{gap1-4},
$\Delta \sigma_y s_x$ and $\Delta \sigma_y s_y$ have point nodes in the
bulk spectrum on the 
the $k_{y}$- and $k_x$-axes, respectively,
while $\Delta \sigma_ys_z$ is full gapped. 
The point nodes change a qualitative structure of
the SABSs as is shown below.
In the following, we focus on $\Delta\sigma_y s_z (\equiv
\hat{\Delta}_{\rm f})$ and $\Delta\sigma_y s_y (\equiv \hat{\Delta}_{\rm n})$
because the result of $\Delta\sigma_y s_x$ is obtained by exchanging
$k_x$ and $k_y$ in that of $\Delta\sigma_y s_y$.

The obtained SABSs in the STI are illustrated in Fig.\ref{edge}.
The SABSs appear when $m_0^2 < \mu^2$.
An important feature of the SABSs is 
that there exists a structural transition of the energy dispersion.
Combined with the nodal structure mentioned above,
the transition results in a variety of Majorana fermions, which we call
(a) cone, (b) caldera, (c) ridge, and (d) valley:
For the full gapped pairing $\hat{\Delta}=\hat{\Delta}_{\rm f}$, we find that
the cone and the caldera are possible. 
For larger values of $\mu$ and $m_1$,
the energy spectrum of the SABS is an axial symmetric and monotonic function
of $k[\equiv(k_x^2+k_y^2)^{1/2}]$, and its shape is a deformed cone
[Fig.\ref{edge}(a)] including higher order terms of $k$. 
For smaller values of $\mu$ and $m_1$, however,
a second crossing of the zero
energy appears at finite $k$ and 
a \textit{caldera} SABS is realized [Fig.\ref{edge}(b)].
This result is consistent with that of Refs.\cite{HaoLee_B11,Hsieh-Fu,hsieh11}.
On the other hand, for the nodal pairing $\hat{\Delta}_{\rm n}$, we obtain the
ridge [Fig.\ref{edge}(c)] and the valley  [Fig.\ref{edge}(d)], instead.  
Although the structural transition occurs on the same critical line,  
Majorana fermions in this case have the flat dispersion
due to the existence of bulk point nodes.
As a result, the cone (caldera) is deformed into the ridge (valley).
We can also show that the flat dispersion 
between the point nodes has a topological origin, thus is
not accidental \cite{YSTY11,STYY11}.

Now we show that the structural transition is intrinsic to the STI.
Due to an argument based on symmetry given below, we find that 
the STI may have a remnant of the surface Dirac fermion in 
the parent topological insulator, and this is the origin of the structural
transition.
To see this, consider how the superconductivity of $\hat{\Delta}_{g=\rm f,n}$ 
may affect on the Dirac fermion.
When $\mu$ is small and in the bottom of the bulk band, the surface
Dirac fermion near the Fermi energy
is well-separated from the bulk band.
Thus, it can be treated apart, and 
the problem reduces to constructing a pairing term of 
the Dirac fermions that is consistent with symmetry of $\hat{\Delta}_g$.
In particular, the induced pairing should have 
odd mirror parity as 
$\hat\Delta_g$,
because the mirror symmetry $\mathcal M_i$ is a
good symmetry on the surface at $z=0$.
However, one find that no pairing term is
allowed to be consistent with the symmetries. 
This means that the Dirac fermion 
remains to be gapless near the Fermi energy when adding $\hat{\Delta}_g$,
in contrast to the case of conventional $s$-wave pairing \cite{Fu1,Sato03}. 
By hybridizing with the Majorana cone (ridge) specific to TSCs,
the gapless Dirac fermion results in a complex caldera (valley)
structure of the Majorana fermions. 
Now consider tuning $\mu$ deep into the bulk band.
As one increases $\mu$, the surface state near the Fermi energy merges
into the bulk band, and finally disappears. 
One now obtains conventional Majorana cone or Majorana ridge because of
no hybridization of the Dirac fermion.
Therefore, a structural transition of the Majorana fermions must occur
between these two limits.

We note that when the transition occurs, the velocity of the Majorana
fermions at $(k_x,k_y)=0$
changes its sign. 
The velocity along the $x$-direction 
$\tilde{v}$ is given by $\tilde v = v a \Delta/{m_0}$ with
\begin{align}
 a = \frac{1 - \sqrt{1 + 4 \tilde m_1 + 4 \tilde m_1^2 \tilde \mu^2}}{2 \tilde m_1\tilde \mu^2},
\label{slope}
\end{align}
where 
$\tilde m_1 = m_0 m_1 / v_z^2$ and  $\tilde \mu = \mu / m_0$.
The transition line determined by $a=0$ is given by $\tilde \mu^2 = 1/(-\tilde m_1)$, which is shown in Fig.\ref{edge}.
Only for the case with $m_0 m_1 < 0$, 
the value of $a$ can become zero, namely, topological insulator triggers the structural transition of SABS.

Now we calculate the tunneling conductance of 
normal metal/STI junction, generalizing theories of 
the tunneling spectroscopy of 
conventional \cite{BTK_B82} and unconventional \cite{KT_R00,TK95}
superconductors. 
We suppose a free electron in the normal metal with  
the Hamiltonian
$H_{N}(\bm k) = [(k_x^2+k_y^2+k_z^2)/(2m_{\rm e}) - \mu_{
N}]\sigma_0 s_0 \tau_z$.
The wave function in the normal metal ($z<0$) is given by
\begin{eqnarray}
 & \hspace{-30em}
 \psi_N (z<0) 
= 
e^{i (k_x x + k_y y)}
\Bigl[
\chi_{\sigma s \rm e} 
e^{i k_{\mathrm ez} z}
\nonumber\\ 
+ \displaystyle\sum_{\sigma' s'} 
\left(
a_{\sigma s \sigma' s'} 
\chi_{\sigma' s' h} 
e^{i k_{\mathrm hz} z}
+
b_{\sigma s \sigma' s'} 
\chi_{\sigma' s' \rm e} 
e^{-i k_{\mathrm ez} z}
\right)\Bigr],
\label{Nstate}
\end{eqnarray}
where $\chi_{\sigma s \tau}$ is the eigenvector of $H_{N}({\bm k})$ with
orbital $\sigma$ and spin $s$ for electron $(\tau = \mathrm e)$ or hole $(\tau
= \mathrm h)$, and $k_{\mathrm ez} = \sqrt{k_{\rm e}^{2} - k^2} = k_{\rm e} \cos \theta$, 
$k_{\rm e} = \sqrt{2m_{\rm e}(\mu_{N} + E)}$, $k_{\mathrm h z}
= \sqrt{2m_{\rm e}(\mu_{N} - E) - k^2 }$, and $a_{\sigma s \sigma' s'}$ ($b_{\sigma s \sigma' s'}$) is the Andreev (normal) reflection coefficient.
The first term of the wave function denotes
an injected electron, and the second (third) one denotes 
a reflected hole (electron)
with reflection coefficient $a_{\sigma s \sigma' s}$ ($b_{\sigma s \sigma' s}$). 
On the other hand, the wave function in the STI side ($z > 0$) is given
by Eq. (\ref{psi}) with the transmission coefficient $t_I$.
These wave functions are connected at the interface ($z = 0$)
by the condition \cite{molenkamp01},
$\psi_{\rm N}(0) =\psi_{\rm STI}(0)$ and 
$v_{\rm N}\psi_{\rm N}(0)=v_{\rm STI}\psi_{\rm STI}(0)$,
with the velocity operator
$v_{\rm N(STI)} = \partial H_{\rm N(STI)} /
\partial k_z|_{k_z \to -i\partial_z}$.  
The above equation determines the
coefficients $a_{\sigma s \sigma' s}$, $b_{\sigma s \sigma' s}$ and $t_{I}$. 
Finally, the normalized charge conductance $G$ is given by
\begin{align}
 \frac{G}{G_{\rm N}}
 = \frac{\displaystyle \sum_{\sigma s} \int_0^{2\pi} d \phi
 \int_0^{\pi/2} d\theta \sin 2\theta \, T_{\sigma s}(\theta,\phi,eV)}
 {\displaystyle \sum_{\sigma s} \int_0^{2\pi} d \phi \int_0^{\pi/2}
 d\theta \sin 2\theta \, T_{\sigma s}(\theta,\phi,0)|_{\Delta=0}},
\end{align}
with the angle resolved transmissivity  
$T_{\sigma s}(\theta,\phi,E) = 1 + \sum_{\sigma' s'} 
(|a_{\sigma s \sigma' s'}|^2 - |b_{\sigma s \sigma' s'}|^2)$
with
$k_x = k\cos\phi$, $k_y= k \sin\phi$, 
where the energy $E$ of the injected electron is fixed at the
bias voltage $eV$. 

In the following, the band mass of the normal metal is fixed as $m_{\rm
e} m_2 = 1$ for simplicity, and
we set $\Delta = 0.6$ meV and 
$\tilde m_1 = -0.59$ or
$\tilde m_1 = -0.17$.
The other parameters are the same as those used in Ref. \cite{SasakiPRL}, i.e.,
$m_0 = -0.28\, \rm eV$, $m_2 = 56.6$ eV\AA$^2$, $v_z=3.09$ eV\AA, $v=4.1$ eV\AA\,    and 
$\tilde \mu = -1.8$.
We control the transmissivity of the normal metal/STI interface by
changing the value of $\mu_{N}$. 
The transmissivity becomes maximum in this model 
for $\mu_{N}/\mu\sim 0.6$  
since the magnitude of 
Fermi momentum in the normal metal coincides with that in STI.
As $\mu_{N}$ increases, the magnitude of 
transmissivity is reduced.

\begin{figure}
\centering
\includegraphics[scale=0.8]{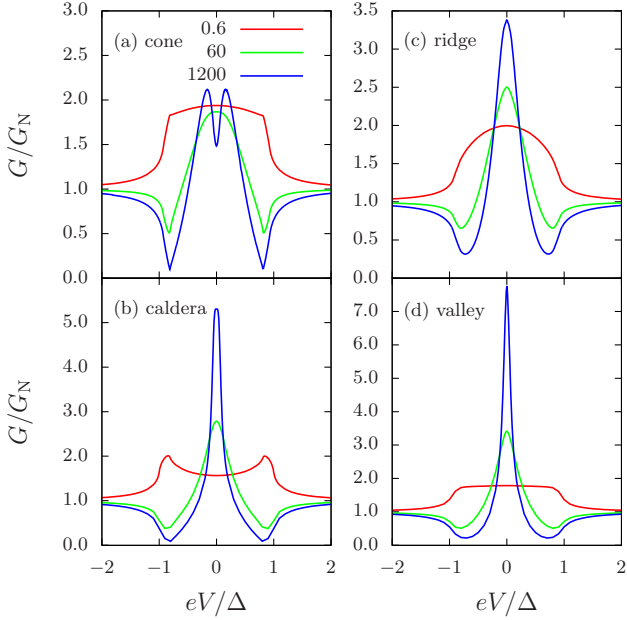}
\caption{(color online) The normalized tunneling conductances $G/G_{N}$ 
near the structural transition of SABSs 
as  functions of bias voltage $eV/\Delta$ for the cone, caldera, ridge and valley SABSs.
The values denoted in the panel are of $\mu/\mu_N$ for each line.
}
\label{conductance}
\end{figure}

The obtained tunneling conductances $G/G_{N}$ near the 
structural transition of SABSs as functions of bias
voltage $eV/\Delta$ are shown in Fig.\ref{conductance}.
With the decrease of the magnitude of transmissivity ($\mu_{N}/\mu=60,
1200$), 
%
the robust ZBCPs appear stemming from the 
gapless SABSs in Fig.\ref{edge}. 
Only for the low transmissivity case with  $\mu_{N}/\mu=1200$ as shown in 
Fig.\ref{conductance} (a), 
$G/G_{N}$ has a double-peak structure. 
The latter is consistent with 
the fact that the corresponding surface local density of states does not have 
a zero energy peak but double peak structure \cite{SasakiPRL}.
In  junctions with high transmissivity with $\mu_{N}/\mu = 0.6$, 
we obtain $G/G_{N}\sim 2$ for $|eV| \sim 0$ which is also 
consistent with the fact that 
an injected electron 
is almost perfectly reflected as a hole due to Andreev 
reflection.  \par

\begin{figure}
\includegraphics[scale=0.8]{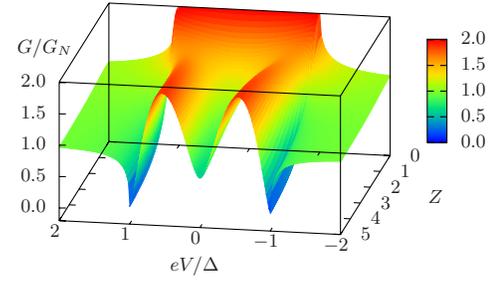}
\caption{
(color online)
The normalized tunneling conductance $G/G_N$ as a function of bias voltage for BW state.}
\label{bw}
\end{figure}

We now focus on STI with $\hat{\Delta}_{\rm f}$.  
It is noted that the difference of the line shapes of 
$G/G_{N}$ 
between Fig.\ref{conductance}(a) and
Fig.\ref{conductance}(b) can be understood from 
the different types of SABSs.  
In the case of Fig.\ref{conductance}(b),
a \textit{caldera}-SABS is realized, 
as shown in Fig.\ref{edge}(b). 
From Eq.(\ref{slope}),
the slope of the dispersion of the SABSs at $k=0$ 
becomes gradual near the structural transition.
This enhances the surface local density of states at $E=0$ and makes the 
ZBCP for the tunneling conductance in the STIs.
%
%
Thus the $G/G_{N}$ at 
$eV=0$ is enhanced in comparison with that for the cone-shaped SABS. 
As a result, 
even in the low transparent limit $\mu_{N}/\mu=1200$, 
no double-peak structure of $G/G_{N}$ appears in 
Fig.\ref{conductance}(b). 
The present feature is different from 
preexisting 3D TSCs 
with spin-triplet $p$-wave pairing realized in BW phase of superfluid $^{3}$He, 
where in contrast to Figs. 2(a) and (b), 
the energy dispersion of the 
SABS becomes a conventional Majorana cone 
\cite{Qi,Schnyder,Nagato,Volovik1,Volovik2,Tsutsumi}.
In this case, the angle resolved transmissivity $T(\theta,\phi,eV)$ is 
given by 
\begin{equation}
T
(\theta,\phi,eV)
=\frac{\sigma_{N}}{2}\sum_{s=\pm1}
\frac{1+ \sigma_{N} | \Gamma |^{2} + (\sigma_{N}-1)|\Gamma|^{4}}
{| 1 + (1 -\sigma_{N}) \Gamma^{2}\exp(-2i\theta s) |^{2}},
\end{equation}
with the transmissivity at the interface $\sigma_{N}$ 
given by $\sigma_{N}=4\cos^{2}\theta /(4\cos^{2}\theta + Z^{2})$ \cite{KT_R00} 
and $\Gamma=\Delta/(eV + \sqrt{(eV)^{2}-\Delta^{2}})$.
$Z$ is a dimensionless constant that controls $\sigma_N$, and 
$\Delta$ is the superconducting gap in this system.
The resulting tunneling conductance  
never shows a ZBCP \cite{Asano}
as shown in Fig.\ref{bw}. 
This difference comes from the absence of the structural transition.
\par
Next, we consider STI with $\hat\Delta_{\rm n}$, 
where the resulting SABS has a quasi-one dimensional 
energy dispersion.  
In the $x$-direction, SABS has a 
flat dispersion as mentioned before [Fig.\ref{edge}(c)].
The present flat dispersion of the SABS makes a ZBCP in $G/G_{N}$ 
for arbitrary lower transmissivity, as shown in
Fig.\ref{conductance}(c). 
When a \textit{valley}-cone is realized as the SABS 
[Fig.\ref{edge}(d)]. 
$G/G_{N}$ at $eV=0$ is enhanced 
[Fig.\ref{conductance}(d)]. 

Finally, we compare our results with the experimentally observed
tunneling spectroscopy in Cu$_x$Bi$_2$Se$_3$.
The tunneling conductance in Au/Ag/Cu$_{0.3}$Bi$_2$Se$_3$ junction has been
observed in Ref.\cite{SasakiPRL}. 
From the lattice constants of Au and Ag ($a\sim 4$\AA) \cite{Ashcroft}, the
Fermi momentum of the normal metal is estimated as 
$k_{F}\sim \pi/a \sim 1 $\AA $^{-1}$, which corresponds to $\mu_{N}/\mu \sim 100$ in
our model.  
While in the actual system, a barrier layer suppressing transmissivity 
could be formed between normal metal and STI, 
it can be taken into account as an effective increase of $\mu_{N}/\mu$.  
Therefore, the experimental result in Ref.\cite{SasakiPRL} should be
compared with ours with
$\mu_{N}/\mu > 100$. 
From Fig.\ref{conductance}, we find that the 
experimentally observed ZBCP is consistent with $\hat\Delta_{\rm f}$ and
$\hat\Delta_{\rm n}$, both of which support ZBCPs 
originating from Majorana fermions on the
normal metal/STI interface. 


In conclusion, we have 
developed a theory of the tunneling spectroscopy of STI.
We have clarified the structural transition of the energy dispersion of
 the SABS, i.e., cone-caldera and ridge-valley transitions, which stems from 
remaining metallic surface states of parent topological insulator.
In the vicinity of the structural transition of SABSs, 
even in the full-gap superconducting 
case, the line shapes of tunneling conductance 
show robust ZBCPs. 
On the other hand, a typical 3D topological superconductor with pair potential 
realized in BW phase in superfluid $^{3}$He, never shows a ZBCP.
Our obtained results serve as a guide to explore  
novel topological superconductors with Majorana fermions 
\cite{Brydon1,Brydon2,Nakosai}. \par

\begin{acknowledgments} 
We thank S. Kawabata, M. Kriener, K. Segawa, S. Sasaki and Y. Ando for useful 
 discussions. 
MS thanks the Kavli Institute for Theoretical Physics, UCSB, for
 hospitality, where this research was completed.
This work was
supported by MEXT (Innovative Area ``Topological
Quantum Phenomena" KAKENHI), and in part by the National Science Foundation under Grant No. NSF PHY05-51164.
\end{acknowledgments}

\bibliography{majorana}

\begin{thebibliography}{57}%
\makeatletter
\providecommand \@ifxundefined [1]{%
 \@ifx{#1\undefined}
}%
\providecommand \@ifnum [1]{%
 \ifnum #1\expandafter \@firstoftwo
 \else \expandafter \@secondoftwo
 \fi
}%
\providecommand \@ifx [1]{%
 \ifx #1\expandafter \@firstoftwo
 \else \expandafter \@secondoftwo
 \fi
}%
\providecommand \natexlab [1]{#1}%
\providecommand \enquote  [1]{``#1''}%
\providecommand \bibnamefont  [1]{#1}%
\providecommand \bibfnamefont [1]{#1}%
\providecommand \citenamefont [1]{#1}%
\providecommand \href@noop [0]{\@secondoftwo}%
\providecommand \href [0]{\begingroup \@sanitize@url \@href}%
\providecommand \@href[1]{\@@startlink{#1}\@@href}%
\providecommand \@@href[1]{\endgroup#1\@@endlink}%
\providecommand \@sanitize@url [0]{\catcode `\\12\catcode `\$12\catcode
  `\&12\catcode `\#12\catcode `\^12\catcode `\_12\catcode `\%12\relax}%
\providecommand \@@startlink[1]{}%
\providecommand \@@endlink[0]{}%
\providecommand \url  [0]{\begingroup\@sanitize@url \@url }%
\providecommand \@url [1]{\endgroup\@href {#1}{\urlprefix }}%
\providecommand \urlprefix  [0]{URL }%
\providecommand \Eprint [0]{\href }%
\providecommand \doibase [0]{http://dx.doi.org/}%
\providecommand \selectlanguage [0]{\@gobble}%
\providecommand \bibinfo  [0]{\@secondoftwo}%
\providecommand \bibfield  [0]{\@secondoftwo}%
\providecommand \translation [1]{[#1]}%
\providecommand \BibitemOpen [0]{}%
\providecommand \bibitemStop [0]{}%
\providecommand \bibitemNoStop [0]{.\EOS\space}%
\providecommand \EOS [0]{\spacefactor3000\relax}%
\providecommand \BibitemShut  [1]{\csname bibitem#1\endcsname}%
\let\auto@bib@innerbib\@empty
\bibitem [{\citenamefont {Tanaka}\ \emph {et~al.}(2012)\citenamefont {Tanaka},
  \citenamefont {Sato},\ and\ \citenamefont {Nagaosa}}]{TSN}%
  \BibitemOpen
  \bibfield  {author} {\bibinfo {author} {\bibfnamefont {Y.}~\bibnamefont
  {Tanaka}}, \bibinfo {author} {\bibfnamefont {M.}~\bibnamefont {Sato}}, \ and\
  \bibinfo {author} {\bibfnamefont {N.}~\bibnamefont {Nagaosa}},\ }\href@noop
  {} {\bibfield  {journal} {\bibinfo  {journal} {J. Phys. Soc. Jpn.}\ }\textbf
  {\bibinfo {volume} {81}},\ \bibinfo {pages} {011013} (\bibinfo {year}
  {2012})}\BibitemShut {NoStop}%
\bibitem [{\citenamefont {Qi}\ and\ \citenamefont {Zhang}(2011)}]{QZ}%
  \BibitemOpen
  \bibfield  {author} {\bibinfo {author} {\bibfnamefont {X.-L.}\ \bibnamefont
  {Qi}}\ and\ \bibinfo {author} {\bibfnamefont {S.-C.}\ \bibnamefont {Zhang}},\
  }\href@noop {} {\bibfield  {journal} {\bibinfo  {journal} {Rev. Mod. Phys.}\
  }\textbf {\bibinfo {volume} {83}},\ \bibinfo {pages} {1057} (\bibinfo {year}
  {2011})}\BibitemShut {NoStop}%
\bibitem [{\citenamefont {Schnyder}\ \emph
  {et~al.}(2008{\natexlab{a}})\citenamefont {Schnyder}, \citenamefont {Ryu},
  \citenamefont {Furusaki},\ and\ \citenamefont {Ludwig}}]{Schny_B08}%
  \BibitemOpen
  \bibfield  {author} {\bibinfo {author} {\bibfnamefont {A.~P.}\ \bibnamefont
  {Schnyder}}, \bibinfo {author} {\bibfnamefont {S.}~\bibnamefont {Ryu}},
  \bibinfo {author} {\bibfnamefont {A.}~\bibnamefont {Furusaki}}, \ and\
  \bibinfo {author} {\bibfnamefont {A.~W.~W.}\ \bibnamefont {Ludwig}},\
  }\href@noop {} {\bibfield  {journal} {\bibinfo  {journal} {Phys. Rev. B}\
  }\textbf {\bibinfo {volume} {78}},\ \bibinfo {pages} {195125} (\bibinfo
  {year} {2008}{\natexlab{a}})}\BibitemShut {NoStop}%
\bibitem [{\citenamefont {Wilczek}(2009)}]{Wil_NP09}%
  \BibitemOpen
  \bibfield  {author} {\bibinfo {author} {\bibfnamefont {F.}~\bibnamefont
  {Wilczek}},\ }\href@noop {} {\bibfield  {journal} {\bibinfo  {journal}
  {Nature Phys.}\ }\textbf {\bibinfo {volume} {5}},\ \bibinfo {pages} {614}
  (\bibinfo {year} {2009})}\BibitemShut {NoStop}%
\bibitem [{\citenamefont {M.Sato}\ and\ \citenamefont
  {S.Fujimoto}(2009)}]{SF09}%
  \BibitemOpen
  \bibfield  {author} {\bibinfo {author} {\bibnamefont {M.Sato}}\ and\ \bibinfo
  {author} {\bibnamefont {S.Fujimoto}},\ }\href@noop {} {\bibfield  {journal}
  {\bibinfo  {journal} {Phys. Rev. B}\ }\textbf {\bibinfo {volume} {79}},\
  \bibinfo {pages} {094504} (\bibinfo {year} {2009})}\BibitemShut {NoStop}%
\bibitem [{\citenamefont {M.Sato}\ \emph {et~al.}(2009)\citenamefont {M.Sato},
  \citenamefont {Takahashi},\ and\ \citenamefont {Fujimoto}}]{STF09}%
  \BibitemOpen
  \bibfield  {author} {\bibinfo {author} {\bibnamefont {M.Sato}}, \bibinfo
  {author} {\bibfnamefont {Y.}~\bibnamefont {Takahashi}}, \ and\ \bibinfo
  {author} {\bibfnamefont {S.}~\bibnamefont {Fujimoto}},\ }\href@noop {}
  {\bibfield  {journal} {\bibinfo  {journal} {Phys. Rev. Lett.}\ }\textbf
  {\bibinfo {volume} {103}},\ \bibinfo {pages} {020401} (\bibinfo {year}
  {2009})}\BibitemShut {NoStop}%
\bibitem [{\citenamefont {M.Sato}\ \emph {et~al.}(2010)\citenamefont {M.Sato},
  \citenamefont {Takahashi},\ and\ \citenamefont {Fujimoto}}]{STF10}%
  \BibitemOpen
  \bibfield  {author} {\bibinfo {author} {\bibnamefont {M.Sato}}, \bibinfo
  {author} {\bibfnamefont {Y.}~\bibnamefont {Takahashi}}, \ and\ \bibinfo
  {author} {\bibfnamefont {S.}~\bibnamefont {Fujimoto}},\ }\href@noop {}
  {\bibfield  {journal} {\bibinfo  {journal} {Phys. Rev. B}\ }\textbf {\bibinfo
  {volume} {82}},\ \bibinfo {pages} {134521} (\bibinfo {year}
  {2010})}\BibitemShut {NoStop}%
\bibitem [{\citenamefont {Sato}\ and\ \citenamefont {Fujimoto}(2010)}]{SF10}%
  \BibitemOpen
  \bibfield  {author} {\bibinfo {author} {\bibfnamefont {M.}~\bibnamefont
  {Sato}}\ and\ \bibinfo {author} {\bibfnamefont {S.}~\bibnamefont
  {Fujimoto}},\ }\href@noop {} {\bibfield  {journal} {\bibinfo  {journal}
  {Phys. Rev. Lett.}\ }\textbf {\bibinfo {volume} {105}},\ \bibinfo {pages}
  {217001} (\bibinfo {year} {2010})}\BibitemShut {NoStop}%
\bibitem [{\citenamefont {J.D.Sau}\ \emph {et~al.}(2010)\citenamefont
  {J.D.Sau}, \citenamefont {R.M.Lutchyn}, \citenamefont {S.Tewari},\ and\
  \citenamefont {Sarma}}]{SLTS10}%
  \BibitemOpen
  \bibfield  {author} {\bibinfo {author} {\bibnamefont {J.D.Sau}}, \bibinfo
  {author} {\bibnamefont {R.M.Lutchyn}}, \bibinfo {author} {\bibnamefont
  {S.Tewari}}, \ and\ \bibinfo {author} {\bibfnamefont {S.~D.}\ \bibnamefont
  {Sarma}},\ }\href@noop {} {\bibfield  {journal} {\bibinfo  {journal} {Phys.
  Rev. Lett.}\ }\textbf {\bibinfo {volume} {104}},\ \bibinfo {pages} {040502}
  (\bibinfo {year} {2010})}\BibitemShut {NoStop}%
\bibitem [{\citenamefont {Alicea}(2010)}]{Alicea10}%
  \BibitemOpen
  \bibfield  {author} {\bibinfo {author} {\bibfnamefont {J.}~\bibnamefont
  {Alicea}},\ }\href@noop {} {\bibfield  {journal} {\bibinfo  {journal} {Phys.
  Rev. B}\ }\textbf {\bibinfo {volume} {81}},\ \bibinfo {pages} {125318}
  (\bibinfo {year} {2010})}\BibitemShut {NoStop}%
\bibitem [{\citenamefont {Lutchyn}\ \emph {et~al.}(2010)\citenamefont
  {Lutchyn}, \citenamefont {Stanescu},\ and\ \citenamefont {Sarma}}]{LSD10}%
  \BibitemOpen
  \bibfield  {author} {\bibinfo {author} {\bibfnamefont {R.~M.}\ \bibnamefont
  {Lutchyn}}, \bibinfo {author} {\bibfnamefont {T.}~\bibnamefont {Stanescu}}, \
  and\ \bibinfo {author} {\bibfnamefont {S.~D.}\ \bibnamefont {Sarma}},\
  }\href@noop {} {\bibfield  {journal} {\bibinfo  {journal} {Phys. Rev. Lett.}\
  }\textbf {\bibinfo {volume} {105}},\ \bibinfo {pages} {077001} (\bibinfo
  {year} {2010})}\BibitemShut {NoStop}%
\bibitem [{\citenamefont {Oreg}\ \emph {et~al.}(2010)\citenamefont {Oreg},
  \citenamefont {Refael},\ and\ \citenamefont {von Oppen}}]{ORO10}%
  \BibitemOpen
  \bibfield  {author} {\bibinfo {author} {\bibfnamefont {Y.}~\bibnamefont
  {Oreg}}, \bibinfo {author} {\bibfnamefont {G.}~\bibnamefont {Refael}}, \ and\
  \bibinfo {author} {\bibfnamefont {F.}~\bibnamefont {von Oppen}},\ }\href@noop
  {} {\bibfield  {journal} {\bibinfo  {journal} {Phys. Rev. Lett.}\ }\textbf
  {\bibinfo {volume} {105}},\ \bibinfo {pages} {177002} (\bibinfo {year}
  {2010})}\BibitemShut {NoStop}%
\bibitem [{\citenamefont {Lutchyn}\ \emph {et~al.}(2011)\citenamefont
  {Lutchyn}, \citenamefont {Stanescu},\ and\ \citenamefont {Sarma}}]{LSS11}%
  \BibitemOpen
  \bibfield  {author} {\bibinfo {author} {\bibfnamefont {R.~M.}\ \bibnamefont
  {Lutchyn}}, \bibinfo {author} {\bibfnamefont {T.}~\bibnamefont {Stanescu}}, \
  and\ \bibinfo {author} {\bibfnamefont {S.~D.}\ \bibnamefont {Sarma}},\
  }\href@noop {} {\bibfield  {journal} {\bibinfo  {journal} {Phys. Rev. Lett.}\
  }\textbf {\bibinfo {volume} {106}},\ \bibinfo {pages} {127001} (\bibinfo
  {year} {2011})}\BibitemShut {NoStop}%
\bibitem [{\citenamefont {Alicea}\ \emph {et~al.}(2011)\citenamefont {Alicea},
  \citenamefont {Oreg}, \citenamefont {Rafael}, \citenamefont {von Oppen},\
  and\ \citenamefont {Fisher}}]{AOROF11}%
  \BibitemOpen
  \bibfield  {author} {\bibinfo {author} {\bibfnamefont {J.}~\bibnamefont
  {Alicea}}, \bibinfo {author} {\bibfnamefont {Y.}~\bibnamefont {Oreg}},
  \bibinfo {author} {\bibfnamefont {G.}~\bibnamefont {Rafael}}, \bibinfo
  {author} {\bibfnamefont {F.}~\bibnamefont {von Oppen}}, \ and\ \bibinfo
  {author} {\bibfnamefont {M.~F.}\ \bibnamefont {Fisher}},\ }\href@noop {}
  {\bibfield  {journal} {\bibinfo  {journal} {Nat. Phys.}\ }\textbf {\bibinfo
  {volume} {7}},\ \bibinfo {pages} {412} (\bibinfo {year} {2011})}\BibitemShut
  {NoStop}%
\bibitem [{\citenamefont {L.Fu}\ and\ \citenamefont {Kane}(2008)}]{Fu1}%
  \BibitemOpen
  \bibfield  {author} {\bibinfo {author} {\bibnamefont {L.Fu}}\ and\ \bibinfo
  {author} {\bibfnamefont {C.~L.}\ \bibnamefont {Kane}},\ }\href@noop {}
  {\bibfield  {journal} {\bibinfo  {journal} {Phys. Rev. Lett.}\ }\textbf
  {\bibinfo {volume} {100}},\ \bibinfo {pages} {096407} (\bibinfo {year}
  {2008})}\BibitemShut {NoStop}%
\bibitem [{\citenamefont {Fu}\ and\ \citenamefont {Kane}(2009)}]{FK09}%
  \BibitemOpen
  \bibfield  {author} {\bibinfo {author} {\bibfnamefont {L.}~\bibnamefont
  {Fu}}\ and\ \bibinfo {author} {\bibfnamefont {C.~L.}\ \bibnamefont {Kane}},\
  }\href@noop {} {\bibfield  {journal} {\bibinfo  {journal} {Phys. Rev. Lett.}\
  }\textbf {\bibinfo {volume} {102}},\ \bibinfo {pages} {216403} (\bibinfo
  {year} {2009})}\BibitemShut {NoStop}%
\bibitem [{\citenamefont {Akhmerov}\ \emph {et~al.}(2009)\citenamefont
  {Akhmerov}, \citenamefont {Nilsson},\ and\ \citenamefont
  {Beenakker}}]{ANB09}%
  \BibitemOpen
  \bibfield  {author} {\bibinfo {author} {\bibfnamefont {A.~R.}\ \bibnamefont
  {Akhmerov}}, \bibinfo {author} {\bibfnamefont {J.}~\bibnamefont {Nilsson}}, \
  and\ \bibinfo {author} {\bibfnamefont {C.~W.~J.}\ \bibnamefont {Beenakker}},\
  }\href@noop {} {\bibfield  {journal} {\bibinfo  {journal} {Phys. Rev. Lett.}\
  }\textbf {\bibinfo {volume} {102}},\ \bibinfo {pages} {216404} (\bibinfo
  {year} {2009})}\BibitemShut {NoStop}%
\bibitem [{\citenamefont {Law}\ \emph {et~al.}(2009)\citenamefont {Law},
  \citenamefont {Lee},\ and\ \citenamefont {Ng}}]{LLN09}%
  \BibitemOpen
  \bibfield  {author} {\bibinfo {author} {\bibfnamefont {K.~T.}\ \bibnamefont
  {Law}}, \bibinfo {author} {\bibfnamefont {P.~A.}\ \bibnamefont {Lee}}, \ and\
  \bibinfo {author} {\bibfnamefont {T.~K.}\ \bibnamefont {Ng}},\ }\href@noop {}
  {\bibfield  {journal} {\bibinfo  {journal} {Phys. Rev. Lett.}\ }\textbf
  {\bibinfo {volume} {103}},\ \bibinfo {pages} {237001} (\bibinfo {year}
  {2009})}\BibitemShut {NoStop}%
\bibitem [{\citenamefont {Tanaka}\ \emph {et~al.}(2009)\citenamefont {Tanaka},
  \citenamefont {Yokoyama},\ and\ \citenamefont {Nagaosa}}]{TYN09}%
  \BibitemOpen
  \bibfield  {author} {\bibinfo {author} {\bibfnamefont {Y.}~\bibnamefont
  {Tanaka}}, \bibinfo {author} {\bibfnamefont {T.}~\bibnamefont {Yokoyama}}, \
  and\ \bibinfo {author} {\bibfnamefont {N.}~\bibnamefont {Nagaosa}},\
  }\href@noop {} {\bibfield  {journal} {\bibinfo  {journal} {Phys. Rev. Lett.}\
  }\textbf {\bibinfo {volume} {103}},\ \bibinfo {pages} {107002} (\bibinfo
  {year} {2009})}\BibitemShut {NoStop}%
\bibitem [{\citenamefont {Linder}\ \emph {et~al.}(2010)\citenamefont {Linder},
  \citenamefont {Tanaka}, \citenamefont {Yokoyama}, \citenamefont {Sudbo},\
  and\ \citenamefont {Nagaosa}}]{Linder10a}%
  \BibitemOpen
  \bibfield  {author} {\bibinfo {author} {\bibfnamefont {J.}~\bibnamefont
  {Linder}}, \bibinfo {author} {\bibfnamefont {Y.}~\bibnamefont {Tanaka}},
  \bibinfo {author} {\bibfnamefont {T.}~\bibnamefont {Yokoyama}}, \bibinfo
  {author} {\bibfnamefont {A.}~\bibnamefont {Sudbo}}, \ and\ \bibinfo {author}
  {\bibfnamefont {N.}~\bibnamefont {Nagaosa}},\ }\href@noop {} {\bibfield
  {journal} {\bibinfo  {journal} {Phys. Rev. Lett.}\ }\textbf {\bibinfo
  {volume} {104}},\ \bibinfo {pages} {067001} (\bibinfo {year}
  {2010})}\BibitemShut {NoStop}%
\bibitem [{\citenamefont {Yamakage}\ \emph {et~al.}(2012)\citenamefont
  {Yamakage}, \citenamefont {Tanaka},\ and\ \citenamefont
  {Nagaosa}}]{yamakage}%
  \BibitemOpen
  \bibfield  {author} {\bibinfo {author} {\bibfnamefont {A.}~\bibnamefont
  {Yamakage}}, \bibinfo {author} {\bibfnamefont {Y.}~\bibnamefont {Tanaka}}, \
  and\ \bibinfo {author} {\bibfnamefont {N.}~\bibnamefont {Nagaosa}},\
  }\href@noop {} {\bibfield  {journal} {\bibinfo  {journal} {Phys. Rev. Lett.}\
  }\textbf {\bibinfo {volume} {108}},\ \bibinfo {pages} {087003} (\bibinfo
  {year} {2012})}\BibitemShut {NoStop}%
\bibitem [{\citenamefont {Hor}\ \emph {et~al.}(2010)\citenamefont {Hor},
  \citenamefont {Williams}, \citenamefont {Checkelsky}, \citenamefont
  {Roushan}, \citenamefont {Seo}, \citenamefont {Xu}, \citenamefont
  {Zandbergen}, \citenamefont {Yazdani}, \citenamefont {Ong},\ and\
  \citenamefont {Cava}}]{Hor_L10}%
  \BibitemOpen
  \bibfield  {author} {\bibinfo {author} {\bibfnamefont {Y.~S.}\ \bibnamefont
  {Hor}}, \bibinfo {author} {\bibfnamefont {A.~J.}\ \bibnamefont {Williams}},
  \bibinfo {author} {\bibfnamefont {J.~G.}\ \bibnamefont {Checkelsky}},
  \bibinfo {author} {\bibfnamefont {P.}~\bibnamefont {Roushan}}, \bibinfo
  {author} {\bibfnamefont {J.}~\bibnamefont {Seo}}, \bibinfo {author}
  {\bibfnamefont {Q.}~\bibnamefont {Xu}}, \bibinfo {author} {\bibfnamefont
  {H.~W.}\ \bibnamefont {Zandbergen}}, \bibinfo {author} {\bibfnamefont
  {A.}~\bibnamefont {Yazdani}}, \bibinfo {author} {\bibfnamefont {N.~P.}\
  \bibnamefont {Ong}}, \ and\ \bibinfo {author} {\bibfnamefont {R.~J.}\
  \bibnamefont {Cava}},\ }\href@noop {} {\bibfield  {journal} {\bibinfo
  {journal} {Phys. Rev. Lett.}\ }\textbf {\bibinfo {volume} {104}},\ \bibinfo
  {pages} {057001} (\bibinfo {year} {2010})}\BibitemShut {NoStop}%
\bibitem [{\citenamefont {Wray}\ \emph {et~al.}(2010)\citenamefont {Wray},
  \citenamefont {Xu}, \citenamefont {Xia}, \citenamefont {Hor}, \citenamefont
  {Qian}, \citenamefont {Fedorov}, \citenamefont {Lin}, \citenamefont {Bansil},
  \citenamefont {Cava},\ and\ \citenamefont {Hasan}}]{Wray_NP10}%
  \BibitemOpen
  \bibfield  {author} {\bibinfo {author} {\bibfnamefont {L.~A.}\ \bibnamefont
  {Wray}}, \bibinfo {author} {\bibfnamefont {S.-Y.}\ \bibnamefont {Xu}},
  \bibinfo {author} {\bibfnamefont {Y.}~\bibnamefont {Xia}}, \bibinfo {author}
  {\bibfnamefont {Y.~S.}\ \bibnamefont {Hor}}, \bibinfo {author} {\bibfnamefont
  {D.}~\bibnamefont {Qian}}, \bibinfo {author} {\bibfnamefont {A.~V.}\
  \bibnamefont {Fedorov}}, \bibinfo {author} {\bibfnamefont {H.}~\bibnamefont
  {Lin}}, \bibinfo {author} {\bibfnamefont {A.}~\bibnamefont {Bansil}},
  \bibinfo {author} {\bibfnamefont {R.~J.}\ \bibnamefont {Cava}}, \ and\
  \bibinfo {author} {\bibfnamefont {M.~Z.}\ \bibnamefont {Hasan}},\ }\href@noop
  {} {\bibfield  {journal} {\bibinfo  {journal} {Nature Phys.}\ }\textbf
  {\bibinfo {volume} {6}},\ \bibinfo {pages} {855} (\bibinfo {year}
  {2010})}\BibitemShut {NoStop}%
\bibitem [{\citenamefont {Kriener}\ \emph
  {et~al.}(2011{\natexlab{a}})\citenamefont {Kriener}, \citenamefont {Segawa},
  \citenamefont {Ren}, \citenamefont {Sasaki},\ and\ \citenamefont
  {Ando}}]{MKR_L11}%
  \BibitemOpen
  \bibfield  {author} {\bibinfo {author} {\bibfnamefont {M.}~\bibnamefont
  {Kriener}}, \bibinfo {author} {\bibfnamefont {K.}~\bibnamefont {Segawa}},
  \bibinfo {author} {\bibfnamefont {Z.}~\bibnamefont {Ren}}, \bibinfo {author}
  {\bibfnamefont {S.}~\bibnamefont {Sasaki}}, \ and\ \bibinfo {author}
  {\bibfnamefont {Y.}~\bibnamefont {Ando}},\ }\href@noop {} {\bibfield
  {journal} {\bibinfo  {journal} {Phys. Rev. Lett.}\ }\textbf {\bibinfo
  {volume} {106}},\ \bibinfo {pages} {127004} (\bibinfo {year}
  {2011}{\natexlab{a}})}\BibitemShut {NoStop}%
\bibitem [{\citenamefont {Kriener}\ \emph
  {et~al.}(2011{\natexlab{b}})\citenamefont {Kriener}, \citenamefont {Segawa},
  \citenamefont {Ren}, \citenamefont {Sasaki}, \citenamefont {Wada},
  \citenamefont {Kuwabata},\ and\ \citenamefont {Ando}}]{MKR2}%
  \BibitemOpen
  \bibfield  {author} {\bibinfo {author} {\bibfnamefont {M.}~\bibnamefont
  {Kriener}}, \bibinfo {author} {\bibfnamefont {K.}~\bibnamefont {Segawa}},
  \bibinfo {author} {\bibfnamefont {Z.}~\bibnamefont {Ren}}, \bibinfo {author}
  {\bibfnamefont {S.}~\bibnamefont {Sasaki}}, \bibinfo {author} {\bibfnamefont
  {S.}~\bibnamefont {Wada}}, \bibinfo {author} {\bibfnamefont {S.}~\bibnamefont
  {Kuwabata}}, \ and\ \bibinfo {author} {\bibfnamefont {Y.}~\bibnamefont
  {Ando}},\ }\href@noop {} {\bibfield  {journal} {\bibinfo  {journal} {Phys.
  Rev. B}\ }\textbf {\bibinfo {volume} {84}},\ \bibinfo {pages} {054513}
  (\bibinfo {year} {2011}{\natexlab{b}})}\BibitemShut {NoStop}%
\bibitem [{\citenamefont {Sato}(2009)}]{Sato_B09}%
  \BibitemOpen
  \bibfield  {author} {\bibinfo {author} {\bibfnamefont {M.}~\bibnamefont
  {Sato}},\ }\href@noop {} {\bibfield  {journal} {\bibinfo  {journal} {Phys.
  Rev. B}\ }\textbf {\bibinfo {volume} {79}},\ \bibinfo {pages} {214526}
  (\bibinfo {year} {2009})}\BibitemShut {NoStop}%
\bibitem [{\citenamefont {Fu}\ and\ \citenamefont {Berg}(2010)}]{FB_L10}%
  \BibitemOpen
  \bibfield  {author} {\bibinfo {author} {\bibfnamefont {L.}~\bibnamefont
  {Fu}}\ and\ \bibinfo {author} {\bibfnamefont {E.}~\bibnamefont {Berg}},\
  }\href@noop {} {\bibfield  {journal} {\bibinfo  {journal} {Phys. Rev. Lett.}\
  }\textbf {\bibinfo {volume} {105}},\ \bibinfo {pages} {097001} (\bibinfo
  {year} {2010})}\BibitemShut {NoStop}%
\bibitem [{\citenamefont {Sato}(2010)}]{Sato_B10}%
  \BibitemOpen
  \bibfield  {author} {\bibinfo {author} {\bibfnamefont {M.}~\bibnamefont
  {Sato}},\ }\href@noop {} {\bibfield  {journal} {\bibinfo  {journal} {Phys.
  Rev. B}\ }\textbf {\bibinfo {volume} {81}},\ \bibinfo {pages} {220504(R)}
  (\bibinfo {year} {2010})}\BibitemShut {NoStop}%
\bibitem [{\citenamefont {Hao}\ and\ \citenamefont {Lee}(2011)}]{HaoLee_B11}%
  \BibitemOpen
  \bibfield  {author} {\bibinfo {author} {\bibfnamefont {L.}~\bibnamefont
  {Hao}}\ and\ \bibinfo {author} {\bibfnamefont {T.~K.}\ \bibnamefont {Lee}},\
  }\href@noop {} {\bibfield  {journal} {\bibinfo  {journal} {Phys. Rev. B}\
  }\textbf {\bibinfo {volume} {83}},\ \bibinfo {pages} {134516} (\bibinfo
  {year} {2011})}\BibitemShut {NoStop}%
\bibitem [{\citenamefont {Sasaki}\ \emph {et~al.}(2011)\citenamefont {Sasaki},
  \citenamefont {Kriener}, \citenamefont {Segawa}, \citenamefont {Yada},
  \citenamefont {Tanaka}, \citenamefont {Sato},\ and\ \citenamefont
  {Ando}}]{SasakiPRL}%
  \BibitemOpen
  \bibfield  {author} {\bibinfo {author} {\bibfnamefont {S.}~\bibnamefont
  {Sasaki}}, \bibinfo {author} {\bibfnamefont {M.}~\bibnamefont {Kriener}},
  \bibinfo {author} {\bibfnamefont {K.}~\bibnamefont {Segawa}}, \bibinfo
  {author} {\bibfnamefont {K.}~\bibnamefont {Yada}}, \bibinfo {author}
  {\bibfnamefont {Y.}~\bibnamefont {Tanaka}}, \bibinfo {author} {\bibfnamefont
  {M.}~\bibnamefont {Sato}}, \ and\ \bibinfo {author} {\bibfnamefont
  {Y.}~\bibnamefont {Ando}},\ }\href@noop {} {\bibfield  {journal} {\bibinfo
  {journal} {Phys. Rev. Lett.}\ }\textbf {\bibinfo {volume} {107}},\ \bibinfo
  {pages} {217001} (\bibinfo {year} {2011})}\BibitemShut {NoStop}%
\bibitem [{\citenamefont {Hsieh}\ and\ \citenamefont {Fu}()}]{Hsieh-Fu}%
  \BibitemOpen
  \bibfield  {author} {\bibinfo {author} {\bibfnamefont {T.~H.}\ \bibnamefont
  {Hsieh}}\ and\ \bibinfo {author} {\bibfnamefont {L.}~\bibnamefont {Fu}},\
  }\href@noop {} {}\Eprint {http://arxiv.org/abs/arXiv:1109.3464}
  {arXiv:1109.3464} \BibitemShut {NoStop}%
\bibitem [{\citenamefont {{Hsieh}}\ and\ \citenamefont {{Fu}}()}]{hsieh11}%
  \BibitemOpen
  \bibfield  {author} {\bibinfo {author} {\bibfnamefont {T.~H.}\ \bibnamefont
  {{Hsieh}}}\ and\ \bibinfo {author} {\bibfnamefont {L.}~\bibnamefont {{Fu}}},\
  }\href@noop {} {}\Eprint {http://arxiv.org/abs/arXiv:1112.1951}
  {arXiv:1112.1951} \BibitemShut {NoStop}%
\bibitem [{\citenamefont {Kirzhner}\ \emph {et~al.}()\citenamefont {Kirzhner},
  \citenamefont {Lahoud}, \citenamefont {Chaska}, \citenamefont {Salman},\ and\
  \citenamefont {Kanigel}}]{Kirzhner}%
  \BibitemOpen
  \bibfield  {author} {\bibinfo {author} {\bibfnamefont {T.}~\bibnamefont
  {Kirzhner}}, \bibinfo {author} {\bibfnamefont {E.}~\bibnamefont {Lahoud}},
  \bibinfo {author} {\bibfnamefont {K.}~\bibnamefont {Chaska}}, \bibinfo
  {author} {\bibfnamefont {Z.}~\bibnamefont {Salman}}, \ and\ \bibinfo {author}
  {\bibfnamefont {A.}~\bibnamefont {Kanigel}},\ }\href@noop {} {}\Eprint
  {http://arxiv.org/abs/arXiv:1111.5805} {arXiv:1111.5805} \BibitemShut
  {NoStop}%
\bibitem [{\citenamefont {Koren}\ \emph {et~al.}()\citenamefont {Koren},
  \citenamefont {Kirzhner}, \citenamefont {Lahoud}, \citenamefont {Chashka},\
  and\ \citenamefont {Kanigel}}]{Koren}%
  \BibitemOpen
  \bibfield  {author} {\bibinfo {author} {\bibfnamefont {G.}~\bibnamefont
  {Koren}}, \bibinfo {author} {\bibfnamefont {T.}~\bibnamefont {Kirzhner}},
  \bibinfo {author} {\bibfnamefont {E.}~\bibnamefont {Lahoud}}, \bibinfo
  {author} {\bibfnamefont {K.~B.}\ \bibnamefont {Chashka}}, \ and\ \bibinfo
  {author} {\bibfnamefont {A.}~\bibnamefont {Kanigel}},\ }\href@noop {}
  {}\Eprint {http://arxiv.org/abs/arXiv:1111.3445} {arXiv:1111.3445}
  \BibitemShut {NoStop}%
\bibitem [{\citenamefont {Yang}\ \emph {et~al.}()\citenamefont {Yang},
  \citenamefont {Ding}, \citenamefont {Qu}, \citenamefont {Shen}, \citenamefont
  {Chen}, \citenamefont {Wei}, \citenamefont {Ji}, \citenamefont {Liu},
  \citenamefont {Fan}, \citenamefont {Yang}, \citenamefont {Xiang}, ,\ and\
  \citenamefont {Lu}}]{FanYang}%
  \BibitemOpen
  \bibfield  {author} {\bibinfo {author} {\bibfnamefont {F.}~\bibnamefont
  {Yang}}, \bibinfo {author} {\bibfnamefont {Y.}~\bibnamefont {Ding}}, \bibinfo
  {author} {\bibfnamefont {F.}~\bibnamefont {Qu}}, \bibinfo {author}
  {\bibfnamefont {J.}~\bibnamefont {Shen}}, \bibinfo {author} {\bibfnamefont
  {J.}~\bibnamefont {Chen}}, \bibinfo {author} {\bibfnamefont {Z.}~\bibnamefont
  {Wei}}, \bibinfo {author} {\bibfnamefont {Z.}~\bibnamefont {Ji}}, \bibinfo
  {author} {\bibfnamefont {G.}~\bibnamefont {Liu}}, \bibinfo {author}
  {\bibfnamefont {J.}~\bibnamefont {Fan}}, \bibinfo {author} {\bibfnamefont
  {C.}~\bibnamefont {Yang}}, \bibinfo {author} {\bibfnamefont {T.}~\bibnamefont
  {Xiang}}, , \ and\ \bibinfo {author} {\bibfnamefont {L.}~\bibnamefont {Lu}},\
  }\href@noop {} {}\Eprint {http://arxiv.org/abs/arXiv:1105.0229}
  {arXiv:1105.0229} \BibitemShut {NoStop}%
\bibitem [{\citenamefont {Tanaka}\ and\ \citenamefont
  {Kashiwaya}(1995)}]{TK95}%
  \BibitemOpen
  \bibfield  {author} {\bibinfo {author} {\bibfnamefont {Y.}~\bibnamefont
  {Tanaka}}\ and\ \bibinfo {author} {\bibfnamefont {S.}~\bibnamefont
  {Kashiwaya}},\ }\href@noop {} {\bibfield  {journal} {\bibinfo  {journal}
  {Phys. Rev. Lett.}\ }\textbf {\bibinfo {volume} {74}},\ \bibinfo {pages}
  {3451} (\bibinfo {year} {1995})}\BibitemShut {NoStop}%
\bibitem [{\citenamefont {Honerkamp}\ and\ \citenamefont
  {Sigrist}(1998)}]{Sigrist}%
  \BibitemOpen
  \bibfield  {author} {\bibinfo {author} {\bibfnamefont {C.}~\bibnamefont
  {Honerkamp}}\ and\ \bibinfo {author} {\bibfnamefont {M.}~\bibnamefont
  {Sigrist}},\ }\href@noop {} {\bibfield  {journal} {\bibinfo  {journal} {J.
  Low Temp. Phys.}\ }\textbf {\bibinfo {volume} {111}},\ \bibinfo {pages} {895}
  (\bibinfo {year} {1998})}\BibitemShut {NoStop}%
\bibitem [{\citenamefont {Yamashiro}\ \emph {et~al.}(1997)\citenamefont
  {Yamashiro}, \citenamefont {Tanaka},\ and\ \citenamefont
  {Kashiwaya}}]{YTK97}%
  \BibitemOpen
  \bibfield  {author} {\bibinfo {author} {\bibfnamefont {M.}~\bibnamefont
  {Yamashiro}}, \bibinfo {author} {\bibfnamefont {Y.}~\bibnamefont {Tanaka}}, \
  and\ \bibinfo {author} {\bibfnamefont {S.}~\bibnamefont {Kashiwaya}},\
  }\href@noop {} {\bibfield  {journal} {\bibinfo  {journal} {Phys. Rev. B}\
  }\textbf {\bibinfo {volume} {56}},\ \bibinfo {pages} {7847} (\bibinfo {year}
  {1997})}\BibitemShut {NoStop}%
\bibitem [{\citenamefont {Kashiwaya}\ \emph {et~al.}(2011)\citenamefont
  {Kashiwaya}, \citenamefont {Kashiwaya}, \citenamefont {Kambara},
  \citenamefont {Furuta}, \citenamefont {Yaguchi}, \citenamefont {Tanaka},\
  and\ \citenamefont {Maeno}}]{Kashiwaya11}%
  \BibitemOpen
  \bibfield  {author} {\bibinfo {author} {\bibfnamefont {S.}~\bibnamefont
  {Kashiwaya}}, \bibinfo {author} {\bibfnamefont {H.}~\bibnamefont
  {Kashiwaya}}, \bibinfo {author} {\bibfnamefont {H.}~\bibnamefont {Kambara}},
  \bibinfo {author} {\bibfnamefont {T.}~\bibnamefont {Furuta}}, \bibinfo
  {author} {\bibfnamefont {H.}~\bibnamefont {Yaguchi}}, \bibinfo {author}
  {\bibfnamefont {Y.}~\bibnamefont {Tanaka}}, \ and\ \bibinfo {author}
  {\bibfnamefont {Y.}~\bibnamefont {Maeno}},\ }\href@noop {} {\bibfield
  {journal} {\bibinfo  {journal} {Phys. Rev. Lett.}\ }\textbf {\bibinfo
  {volume} {107}},\ \bibinfo {pages} {077003} (\bibinfo {year}
  {2011})}\BibitemShut {NoStop}%
\bibitem [{\citenamefont {Asano}\ \emph {et~al.}(2003)\citenamefont {Asano},
  \citenamefont {Tanaka}, \citenamefont {Matsuda},\ and\ \citenamefont
  {Kashiwaya}}]{Asano}%
  \BibitemOpen
  \bibfield  {author} {\bibinfo {author} {\bibfnamefont {Y.}~\bibnamefont
  {Asano}}, \bibinfo {author} {\bibfnamefont {Y.}~\bibnamefont {Tanaka}},
  \bibinfo {author} {\bibfnamefont {Y.}~\bibnamefont {Matsuda}}, \ and\
  \bibinfo {author} {\bibfnamefont {S.}~\bibnamefont {Kashiwaya}},\ }\href@noop
  {} {\bibfield  {journal} {\bibinfo  {journal} {Phys. Rev. B}\ }\textbf
  {\bibinfo {volume} {68}},\ \bibinfo {pages} {184506} (\bibinfo {year}
  {2003})}\BibitemShut {NoStop}%
\bibitem [{\citenamefont {Qi}\ \emph {et~al.}(2009)\citenamefont {Qi},
  \citenamefont {Hughes}, \citenamefont {Raghu},\ and\ \citenamefont
  {Zhang}}]{Qi}%
  \BibitemOpen
  \bibfield  {author} {\bibinfo {author} {\bibfnamefont {X.~L.}\ \bibnamefont
  {Qi}}, \bibinfo {author} {\bibfnamefont {T.~L.}\ \bibnamefont {Hughes}},
  \bibinfo {author} {\bibfnamefont {S.}~\bibnamefont {Raghu}}, \ and\ \bibinfo
  {author} {\bibfnamefont {S.~C.}\ \bibnamefont {Zhang}},\ }\href@noop {}
  {\bibfield  {journal} {\bibinfo  {journal} {Phys. Rev. Lett.}\ }\textbf
  {\bibinfo {volume} {102}},\ \bibinfo {pages} {187001} (\bibinfo {year}
  {2009})}\BibitemShut {NoStop}%
\bibitem [{\citenamefont {Schnyder}\ \emph
  {et~al.}(2008{\natexlab{b}})\citenamefont {Schnyder}, \citenamefont {Ryu},
  \citenamefont {Furusaki},\ and\ \citenamefont {Ludwig}}]{Schnyder}%
  \BibitemOpen
  \bibfield  {author} {\bibinfo {author} {\bibfnamefont {A.~P.}\ \bibnamefont
  {Schnyder}}, \bibinfo {author} {\bibfnamefont {S.}~\bibnamefont {Ryu}},
  \bibinfo {author} {\bibfnamefont {A.}~\bibnamefont {Furusaki}}, \ and\
  \bibinfo {author} {\bibfnamefont {A.~W.~W.}\ \bibnamefont {Ludwig}},\
  }\href@noop {} {\bibfield  {journal} {\bibinfo  {journal} {Phys. Rev. B}\
  }\textbf {\bibinfo {volume} {78}},\ \bibinfo {pages} {195125} (\bibinfo
  {year} {2008}{\natexlab{b}})}\BibitemShut {NoStop}%
\bibitem [{\citenamefont {Nagato}\ \emph {et~al.}(2009)\citenamefont {Nagato},
  \citenamefont {Higashitani},\ and\ \citenamefont {Nagai}}]{Nagato}%
  \BibitemOpen
  \bibfield  {author} {\bibinfo {author} {\bibfnamefont {Y.}~\bibnamefont
  {Nagato}}, \bibinfo {author} {\bibfnamefont {S.}~\bibnamefont {Higashitani}},
  \ and\ \bibinfo {author} {\bibfnamefont {K.}~\bibnamefont {Nagai}},\
  }\href@noop {} {\bibfield  {journal} {\bibinfo  {journal} {J. Phys. Soc.
  Jpn.}\ }\textbf {\bibinfo {volume} {78}},\ \bibinfo {pages} {123603}
  (\bibinfo {year} {2009})}\BibitemShut {NoStop}%
\bibitem [{\citenamefont {Volovik}(2009{\natexlab{a}})}]{Volovik1}%
  \BibitemOpen
  \bibfield  {author} {\bibinfo {author} {\bibfnamefont {G.}~\bibnamefont
  {Volovik}},\ }\href@noop {} {\bibfield  {journal} {\bibinfo  {journal} {JETP
  Lett.}\ }\textbf {\bibinfo {volume} {90}},\ \bibinfo {pages} {587} (\bibinfo
  {year} {2009}{\natexlab{a}})}\BibitemShut {NoStop}%
\bibitem [{\citenamefont {Volovik}(2009{\natexlab{b}})}]{Volovik2}%
  \BibitemOpen
  \bibfield  {author} {\bibinfo {author} {\bibfnamefont {G.}~\bibnamefont
  {Volovik}},\ }\href@noop {} {\bibfield  {journal} {\bibinfo  {journal} {JETP
  Lett.}\ }\textbf {\bibinfo {volume} {90}},\ \bibinfo {pages} {398} (\bibinfo
  {year} {2009}{\natexlab{b}})}\BibitemShut {NoStop}%
\bibitem [{\citenamefont {Tsutsumi}\ \emph {et~al.}(2011)\citenamefont
  {Tsutsumi}, \citenamefont {Ichioka},\ and\ \citenamefont
  {Machida}}]{Tsutsumi}%
  \BibitemOpen
  \bibfield  {author} {\bibinfo {author} {\bibfnamefont {Y.}~\bibnamefont
  {Tsutsumi}}, \bibinfo {author} {\bibfnamefont {M.}~\bibnamefont {Ichioka}}, \
  and\ \bibinfo {author} {\bibfnamefont {K.}~\bibnamefont {Machida}},\
  }\href@noop {} {\bibfield  {journal} {\bibinfo  {journal} {Phys. Rev. B}\
  }\textbf {\bibinfo {volume} {83}},\ \bibinfo {pages} {094510} (\bibinfo
  {year} {2011})}\BibitemShut {NoStop}%
\bibitem [{\citenamefont {Fu}(2009)}]{fu09b}%
  \BibitemOpen
  \bibfield  {author} {\bibinfo {author} {\bibfnamefont {L.}~\bibnamefont
  {Fu}},\ }\href@noop {} {\bibfield  {journal} {\bibinfo  {journal} {Phys. Rev.
  Lett.}\ }\textbf {\bibinfo {volume} {103}},\ \bibinfo {pages} {266801}
  (\bibinfo {year} {2009})}\BibitemShut {NoStop}%
\bibitem [{\citenamefont {Yada}\ \emph {et~al.}(2011)\citenamefont {Yada},
  \citenamefont {Sato}, \citenamefont {Tanaka},\ and\ \citenamefont
  {Yokoyama}}]{YSTY11}%
  \BibitemOpen
  \bibfield  {author} {\bibinfo {author} {\bibfnamefont {K.}~\bibnamefont
  {Yada}}, \bibinfo {author} {\bibfnamefont {M.}~\bibnamefont {Sato}}, \bibinfo
  {author} {\bibfnamefont {Y.}~\bibnamefont {Tanaka}}, \ and\ \bibinfo {author}
  {\bibfnamefont {T.}~\bibnamefont {Yokoyama}},\ }\href@noop {} {\bibfield
  {journal} {\bibinfo  {journal} {Phys. Rev. B}\ }\textbf {\bibinfo {volume}
  {83}},\ \bibinfo {pages} {064505} (\bibinfo {year} {2011})}\BibitemShut
  {NoStop}%
\bibitem [{\citenamefont {Sato}\ \emph {et~al.}(2011)\citenamefont {Sato},
  \citenamefont {Tanaka}, \citenamefont {Yada},\ and\ \citenamefont
  {Yokoyama}}]{STYY11}%
  \BibitemOpen
  \bibfield  {author} {\bibinfo {author} {\bibfnamefont {M.}~\bibnamefont
  {Sato}}, \bibinfo {author} {\bibfnamefont {Y.}~\bibnamefont {Tanaka}},
  \bibinfo {author} {\bibfnamefont {K.}~\bibnamefont {Yada}}, \ and\ \bibinfo
  {author} {\bibfnamefont {T.}~\bibnamefont {Yokoyama}},\ }\href@noop {}
  {\bibfield  {journal} {\bibinfo  {journal} {Phys. Rev. B}\ }\textbf {\bibinfo
  {volume} {83}},\ \bibinfo {pages} {224511} (\bibinfo {year}
  {2011})}\BibitemShut {NoStop}%
\bibitem [{\citenamefont {M.Sato}(2003)}]{Sato03}%
  \BibitemOpen
  \bibfield  {author} {\bibinfo {author} {\bibnamefont {M.Sato}},\ }\href@noop
  {} {\bibfield  {journal} {\bibinfo  {journal} {Phys. Lett. B}\ }\textbf
  {\bibinfo {volume} {575}},\ \bibinfo {pages} {126} (\bibinfo {year}
  {2003})}\BibitemShut {NoStop}%
\bibitem [{\citenamefont {Blonder}\ \emph {et~al.}(1982)\citenamefont
  {Blonder}, \citenamefont {Tinkham},\ and\ \citenamefont
  {Klapwijk}}]{BTK_B82}%
  \BibitemOpen
  \bibfield  {author} {\bibinfo {author} {\bibfnamefont {G.~E.}\ \bibnamefont
  {Blonder}}, \bibinfo {author} {\bibfnamefont {M.}~\bibnamefont {Tinkham}}, \
  and\ \bibinfo {author} {\bibfnamefont {T.~M.}\ \bibnamefont {Klapwijk}},\
  }\href@noop {} {\bibfield  {journal} {\bibinfo  {journal} {Phys. Rev. B}\
  }\textbf {\bibinfo {volume} {25}},\ \bibinfo {pages} {4515} (\bibinfo {year}
  {1982})}\BibitemShut {NoStop}%
\bibitem [{\citenamefont {Kashiwaya}\ and\ \citenamefont
  {Tanaka}(2000)}]{KT_R00}%
  \BibitemOpen
  \bibfield  {author} {\bibinfo {author} {\bibfnamefont {S.}~\bibnamefont
  {Kashiwaya}}\ and\ \bibinfo {author} {\bibfnamefont {Y.}~\bibnamefont
  {Tanaka}},\ }\href@noop {} {\bibfield  {journal} {\bibinfo  {journal} {Rep.
  Prg. Phys.}\ }\textbf {\bibinfo {volume} {63}},\ \bibinfo {pages} {1641}
  (\bibinfo {year} {2000})}\BibitemShut {NoStop}%
\bibitem [{\citenamefont {Molenkamp}\ \emph {et~al.}(2001)\citenamefont
  {Molenkamp}, \citenamefont {Schmidt},\ and\ \citenamefont
  {Bauer}}]{molenkamp01}%
  \BibitemOpen
  \bibfield  {author} {\bibinfo {author} {\bibfnamefont {L.~W.}\ \bibnamefont
  {Molenkamp}}, \bibinfo {author} {\bibfnamefont {G.}~\bibnamefont {Schmidt}},
  \ and\ \bibinfo {author} {\bibfnamefont {G.~E.~W.}\ \bibnamefont {Bauer}},\
  }\href {\doibase 10.1103/PhysRevB.64.121202} {\bibfield  {journal} {\bibinfo
  {journal} {Phys. Rev. B}\ }\textbf {\bibinfo {volume} {64}},\ \bibinfo
  {pages} {121202} (\bibinfo {year} {2001})}\BibitemShut {NoStop}%
\bibitem [{\citenamefont {Ashcroft}\ and\ \citenamefont
  {Mermin}(1976)}]{Ashcroft}%
  \BibitemOpen
  \bibfield  {author} {\bibinfo {author} {\bibfnamefont {N.~W.}\ \bibnamefont
  {Ashcroft}}\ and\ \bibinfo {author} {\bibfnamefont {N.~D.}\ \bibnamefont
  {Mermin}},\ }\href@noop {} {\emph {\bibinfo {title} {{Solid State
  Physics}}}}\ (\bibinfo  {publisher} {Saunders College},\ \bibinfo {address}
  {Philadelphia},\ \bibinfo {year} {1976})\BibitemShut {NoStop}%
\bibitem [{\citenamefont {Schnyder}\ \emph {et~al.}(2010)\citenamefont
  {Schnyder}, \citenamefont {Brydon}, \citenamefont {Manske},\ and\
  \citenamefont {Timm}}]{Brydon1}%
  \BibitemOpen
  \bibfield  {author} {\bibinfo {author} {\bibfnamefont {A.~P.}\ \bibnamefont
  {Schnyder}}, \bibinfo {author} {\bibfnamefont {P.~M.~R.}\ \bibnamefont
  {Brydon}}, \bibinfo {author} {\bibfnamefont {D.}~\bibnamefont {Manske}}, \
  and\ \bibinfo {author} {\bibfnamefont {C.}~\bibnamefont {Timm}},\ }\href@noop
  {} {\bibfield  {journal} {\bibinfo  {journal} {Phys. Rev. B}\ }\textbf
  {\bibinfo {volume} {82}},\ \bibinfo {pages} {184508} (\bibinfo {year}
  {2010})}\BibitemShut {NoStop}%
\bibitem [{\citenamefont {Schnyder}\ \emph {et~al.}(2012)\citenamefont
  {Schnyder}, \citenamefont {Brydon},\ and\ \citenamefont {Timm}}]{Brydon2}%
  \BibitemOpen
  \bibfield  {author} {\bibinfo {author} {\bibfnamefont {A.~P.}\ \bibnamefont
  {Schnyder}}, \bibinfo {author} {\bibfnamefont {P.~M.~R.}\ \bibnamefont
  {Brydon}}, \ and\ \bibinfo {author} {\bibfnamefont {C.}~\bibnamefont
  {Timm}},\ }\href@noop {} {\bibfield  {journal} {\bibinfo  {journal} {Phys.
  Rev. B}\ }\textbf {\bibinfo {volume} {85}},\ \bibinfo {pages} {024522}
  (\bibinfo {year} {2012})}\BibitemShut {NoStop}%
\bibitem [{\citenamefont {Nakosai}\ \emph {et~al.}()\citenamefont {Nakosai},
  \citenamefont {Tanaka},\ and\ \citenamefont {Nagaosa}}]{Nakosai}%
  \BibitemOpen
  \bibfield  {author} {\bibinfo {author} {\bibfnamefont {S.}~\bibnamefont
  {Nakosai}}, \bibinfo {author} {\bibfnamefont {Y.}~\bibnamefont {Tanaka}}, \
  and\ \bibinfo {author} {\bibfnamefont {N.}~\bibnamefont {Nagaosa}},\
  }\href@noop {} {}\Eprint {http://arxiv.org/abs/arXiv:1112.5822}
  {arXiv:1112.5822} \BibitemShut {NoStop}%
\end{thebibliography}%

\end{document}